\documentclass[prl,aps,twocolumn,epsfig,superscriptaddress,showpacs]{revtex4}
\usepackage{amsfonts}
\usepackage[dvips]{graphicx}

\usepackage{mathrsfs}
\usepackage[intlimits]{amsmath}

\begin{document}
\author{Jing-Ling Chen}
 \email{chenjl@nankai.edu.cn}
\affiliation{Theoretical Physics Division, Chern Institute of
Mathematics, Nankai University, Tianjin 300071, People's Republic of
China}
\author{Dong-Ling Deng}
 \affiliation{Theoretical Physics Division, Chern Institute of
Mathematics, Nankai University, Tianjin 300071, People's Republic of
China}

\date{\today}

\title{Tight Correlation-Function Bell Inequality for Multipartite $d$-Dimensional System}

\begin{abstract}
We generalize the correlation functions of the
Clauser-Horne-Shimony-Holt (CHSH) inequality to multipartite
$d$-dimensional systems. All the Bell inequalities based on this
generalization take the same simple form as the CHSH inequality. For
small systems, numerical results show that the new inequalities are
tight and we believe this is also valid for higher dimensional
systems. Moreover, the new inequalities are relevant to the previous
ones and for bipartite system, our inequality is equivalent to the
Collins-Gisin-Linden-Masser-Popescu (CGLMP) inequality.
\end{abstract}

 \maketitle

That local and realistic theories impose certain constraints in the
form of some inequalities on statistical correlations of
measurements on multiparticles was first shown by Bell in
$1964$~\cite{J.S.Bell}. Bell pointed out that any kind of local
hidden variable theory should obey these inequalities, while they
can be violated easily in quantum mechanics. After Bell's
applaudable progress, extensive works on Bell inequalities have been
done, including both theoretical analysis and experiment test. For
instance, the Clauser-Horne-Shimony-Holt (CHSH)~\cite{J.Clauser}
inequality was proposed in $1969$, which is more convenient for
experiment to test the non-locality of two $2$-dimensional (qubit)
system.
However, there exists a long-living open question: ``What are the
general inequalities for more complicated situations?" i.e., for
more particles and higher-dimensional systems.

On the one hand, for higher dimensions of two particles, Collins
{\it et al.} constructed a CHSH type inequality for arbitrary
$d$-dimensional (qudit) systems  in $2002$, now known as the
Collins-Gisin-Linden-Masser-Popescu (CGLMP)
inequality~\cite{2002CGLMP-inequality}. This inequality was shown to
be tight, i.e., it defines one of the facets of the convex
polytope~\cite{A.Peres-1998} of local-realistic (LR)
models~\cite{L.Masanes}. There are some other alternative forms of
this inequality~\cite{Li-Bin Fu, 2008-Zohren-Gill}, and the maximal
quantum violation of this inequality was analyzed in
\cite{Chen-max-2qudits-violation}, which showed that the maximal
violation of this inequality occurs at the non-maximally entangled
state. More recently, Seung Woo Lee and Dieter Jaksch introduced
another tight Bell inequality which is maximally violated by
maximally entangled states~\cite{Seung-Woo Lee and Dieter Jaksch}.

On the other hand, there are also various Bell inequalities for $N$
$(N>2)$ particles. In $1990$, Mermin, in the first time, produced a
series of two setting inequalities for arbitrary many
qubits~\cite{N.D.Mermin}. A complementary series of inequalities was
introduced by Ardehali~\cite{M.Ardehali}. In the next step,
Belinskii and Klyshko gave a series of two setting inequalities,
which contained the tight inequalities of Mermin and
Ardehali~\cite{A.V.Belinskii-D.N.Klyshko}. These inequalities, now
known as Mermin-Ardehali-Belinskii-Klyshko (MABK) inequalities, are
maximally violated by the Greenberger-Horne-Zeilinger (GHZ) states
$|\psi\rangle_{GHZ}=\frac{1}{\sqrt{2}}(|0\cdots0\rangle+|1\cdots1\rangle$.
But for the generalized GHZ states
$|\psi'\rangle_{GHZ}=\cos\xi|0\cdots0\rangle+\sin\xi|1\cdots\rangle$
and $N$ odd, there exists one region
$\xi\in(0,\frac{1}{2}\arcsin(1/\sqrt{2^{N-1}})]$ in which the MABK
inequalities are not
violated\cite{V.Scarani-N.Gisin-2001,M.Zukowski-2002}. Thus  the
MABK inequalities may not be the `natural' generalizations of the
CHSH inequality to more than two qubits,  in the sense that the CHSH
inequality violates all the pure states of two-qubit systems. In
$2004$, Chen \textit{et al.} presented a two-setting Bell inequality
for three qubits which can be seen numerically to be violated by any
pure entangled state~\cite{Chen-3qubits-2004}.
In~\cite{Chen-3qudits-BI}, tight Bell inequalities for three
particles with low dimension are presented. Nevertheless, up to now,
there is no generic tight Bell inequality for arbitrary $N$-qudit
systems, even for three qudits, no such inequality has been found.
Since many of quantum communication schemes, such as multiparty key
(secret) sharing~\cite{V.Scarani-2001} and quantum communication
complexity problems~\cite{C. Brukner-2004}, can be measured with
multiparty Bell inequalities of some form~\cite{A. Acin-2004},
derivations of multiparty Bell inequalities are thus one of the most
important and challenging subject in quantum theory.

The purpose of this paper is to present  general  Bell inequalities
based on the correlation functions for $N$-qudit systems. These
inequalities, obtained  by using the same method in \cite{Li-Bin
Fu}, are tight and relevant to the previous ones.
What's more, all the Bell inequalities based on this generalization
take the same simple form as the CHSH inequality. For bipartite
systems, our inequality is equivalent to the (CGLMP) inequality.
However, to be honest, there are two disadvantages for these new
inequalities: (i) The quantum violations of these inequalities are
small and they are not as strong as the previous inequalities,
namely, they are less resistant to noise; (ii) Some pure states do
not violate these inequalities. It is interesting to note that these
two disadvantages indicate that a tight Bell inequality may not
always be the optimal one.

The approach to our new tight Bell inequalities for $N$-qudit
systems is based on the Gedanken experiment. Consider $N$ spatially
separated parties and allow each of them to choose independently
between two dichotomic observables. Let $X^{[1]}_j$, $X^{[2]}_j$
$(j=1, 2, \cdots, N)$ denote the two observables on the $j$th party,
each of them have $d$ possible outcomes: $x^{[1]}_j,
x^{[2]}_j=0,1,\cdots,d-1$ $(j=1,2,\cdots,N)$. The joint
probabilities are denoted by $P(X_1^{[i_1]},\cdots, X_j^{[i_j]})$,
which should satisfy the normalization condition:
\begin{eqnarray}
\sum_{x_1^{[i_1]},\cdots,
x_j^{[i_j]}=0}^{d-1}P(X_1^{[i_1]}=x_1^{[i_1]},\cdots,X_j^{[i_j]}=x_j^{[i_j]})=1.
\end{eqnarray}

For two-qudit systems, namely $N=2$, Ref.~\cite{Li-Bin Fu}
introduced the correlation functions $Q_{ij}$ in the following form:
\begin{eqnarray}
Q_{ij}=\frac{1}{S}\sum_{m=0}^{d-1}\sum_{n=0}^{d-1}f^{ij}(m,n)P(X_1^{[i]}=m,X_2^{[j]}=n),
\end{eqnarray}
where $S=(d-1)/2$ is the spin of the particle for the
$d$-dimensional system. $f^{ij}(m,n)=S-M[\varepsilon(i-j)(m+n),d]$;
$\varepsilon(x)=1$ and $-1$ for $x\geq0$ and $x<0$, respectively;
$M(x,d)=(x$ mod  $d)$ and $0\leq M(x,d)\leq d-1$. Based on this
correlation functions, a tight Bell inequality for two qudits is
generalized as:
\begin{eqnarray}\label{corr-2qudits-BI}
I_d^{[2]}=Q_{11}+Q_{12}-Q_{21}+Q_{22}\leq2.
\end{eqnarray}
Inequality (\ref{corr-2qudits-BI}) is equivalent to the CGLMP
inequality and its maximal quantum violation is analyzed in
Ref.~\cite{2008-Zohren-Gill,Chen-max-2qudits-violation}.

Inspired by the ideas in Ref.~\cite{Li-Bin Fu}, we generalized the
correlation functions for multipartite $d$-dimensional systems. For
simplicity and convenience, we will focus on the three-qudit case at
first and then analyze the general case. For three-qudit systems,
namely $N=3$, the new correlation functions  $Q_{ijk}$ can be
written in the following form:
\begin{widetext}
\begin{eqnarray}\label{Corr function}
Q_{ijk}\equiv\frac{1}{S}\sum_{m=0}^{d-1}\sum_{n=0}^{d-1}\sum_{l=0}^{d-1}f^{ijk}(m,n,l)
P(X_1^{[i]}=m,X_2^{[j]}=n,X_3^{[k]}=l),
\end{eqnarray}
\end{widetext}
where $S$ takes the same value as the two qudits case: $S=(d-1)/2$,
and $f^{ijk}(m,n,l)=S-M[(-1)^{i\times j \times k}(m+n+l),d]$. Then
the Bell inequality for three particles $d$-dimensional systems
reads:
\begin{eqnarray}\label{three-BI}
I^{[3]}_d=Q_{111}-Q_{222}+Q_{121}+Q_{212}\leq 2.
\end{eqnarray}
Obviously, $I_d^{[3]}$ is upper bounded by $4$ since the extreme
values of $Q_{ijk}$ are $\pm1$ and it can never reach this value
because that the four functions in Eq.(\ref{three-BI}) are strongly
correlated.
In fact, for local hidden variable theories, it is easy to prove
that the maximum value of $I^{[3]}_d$ is $2$.
We use the same method as for two $d$-dimensional systems in
Ref.~\cite{Li-Bin Fu}. The essential idea of this proof is to
enumerate all the possible relations between pairs of operators.
Defining $r_{111}\equiv X_1^{[1]}+X_2^{[1]}+X_3^{[1]}$,
$r_{222}\equiv X_1^{[2]}+X_2^{[2]}+X_3^{[2]}$, $r_{121}\equiv
X_1^{[1]}+X_2^{[2]}+X_3^{[1]}$, and $r_{212}\equiv
X_1^{[2]}+X_2^{[1]}+X_3^{[2]}$. Then the constraint follows
immediately:
\begin{eqnarray}\label{ralations of r}
r_{111}+r_{222}=r_{121}+r_{212}.
\end{eqnarray}
For convenience, we define two functions:
$g_1(x)=\frac{S-M(x,d)}{S}, g_2(x)=\frac{M(x,d)-S-1}{S}$.
Then, for a given choice of $r_{111}$, $r_{222}$,  $r_{121}$, and
$r_{212}$, the correlation functions in Eq.(\ref{three-BI}) can be
rewritten as: $Q_{111}=g_2(r_{111})$, $Q_{222}=g_1(r_{222})$,
$Q_{121}=g_1(r_{121})$, and $Q_{212}=g_1(r_{212})$. A direct
calculation shows that:
\begin{eqnarray}
I^{[3]}_d=\frac{1}{S}&[&M(r_{111},d)+M(r_{222},d)\nonumber\\
&&-M(r_{121},d)- M(r_{212},d)-1].
\end{eqnarray}
Now, we should enumerate all the possible cases according to the
different values of $r_{111}$,,$r_{222}$, $r_{121}$, and $r_{212}$.

Case $1$: Both $r_{111}$ and $r_{222}$ are less than $d$. From
(\ref{ralations of r}), there are two cases for the rest:(i) none of
$r_{212}$ and $r_{121}$ is larger than $d$. then we have
$I_d^{[3]}=[r_{111}+r_{222}-(r_{212}+r_{121})-1]/S=-1/S$ (note that
$d=2S+1$); (ii) one of $r_{212}$ and $r_{121}$ is equal to or larger
than $d$. Then after some simple calculating, we get
$I_{d}^{[3]}=(d-1)/S=2$.

Case $2$: $r_{111}<d$ and $d\leq r_{222}<2d$ or $d\leq r_{111}<2d$
and $r_{222}<d$. From (\ref{ralations of r}), there are four cases
for the rest: (i) both $r_{212}$ and $r_{121}$ are less than $d$.
then we have
$I_d^{[3]}=[r_{111}+r_{222}-d-(r_{212}+r_{121})-1]/S=-2(S+1)/S$;
(ii) one of $r_{212}$ and $r_{121}$ is equal to or larger than $d$.
Then after some simple calculating, we get $I_{d}^{[3]}=-1/S$;(iii)
Both $r_{212}$ and $r_{121}$ are larger than $d$ and less than $2d$,
then $I_d^{[3]}=2$; (iv) one of $r_{212}$ and $r_{121}$ is less than
$d$, and the other is larger than $2d$, then we can also get
$I_d^{[3]}=2$.

Case $3$: $d\leq r_{111}<2d$ and $d\leq r_{222}<2d$. From
(\ref{ralations of r}), there are four cases for the rest: (i) one
of $r_{212}$ and $r_{121}$ is less than $d$ and the other is larger
than $d$ and less than $2d$. then we have $I_d^{[3]}=-2(S+1)/S$;
(ii) both of them are larger than $d$ and less than $2d$. Then
obviously, $I_{d}^{[3]}=-1/S$; (iii) one of them is larger than $2d$
and the other is less than $d$, then $I_d^{[3]}=-1/s$; (iv) one of
them is larger than $2d$ and the other is larger than $d$ and less
than $2d$, then $I_d^{[3]}=2$.

Case $4$: Both $r_{111}$ and $r_{222}$ are equal to or larger than
$2d$ . From (\ref{ralations of r}), there are two cases  for the
rest:(i) one of $r_{212}$ and $r_{121}$ is larger than $2d$ and the
other is larger than $d$ and less than $2d$. then we have
$I_d=-2(S+1)/S$; (ii) both of them are larger than $2d$, then
obviously, $I_{d}^{[3]}=-1/S$.

Thus, we have proved that $I_d^{[3]}\leq2$ for local realistic
theories (Note that for $d=2$, $I_2^{[3]}$ has only two possible
values $\pm2$ since not all the possibilities enumerated above can
occur). Moreover, we have found computationally that the
inequality~(\ref{three-BI}) is tight for $d\leq10$, and suspect that
this will generalize.
If we set $X_3^{[1]}=0$ and $X_3^{[2]}=0$, then the inequality
(\ref{three-BI}) reduces to a two qudits Bell inequality which is an
alternative form of inequality~(\ref{corr-2qudits-BI}) and
equivalent to the CGLMP inequality.

Let us now focus on the quantum violation of the inequality
~(\ref{three-BI}). We will restrict the considerations to multi-port
beamsplitters since the software takes too long to run on our
computer if the most general measurements are employed. Actually,
for low dimensional systems ($d\leq3$), we have used the most
general measurements but no larger violations are founded.
In a Gedanken experiment~\cite{M. Zukowski-Z.Zeilinger}, the matrix
elements of an unbiased symmetric multi-port beamsplitter are given
by
$U_{kl}(\vec{\varphi})=\frac{1}{\sqrt{d}}\alpha^{kl}$exp$(i\varphi^l)$,
here $\alpha=$exp$(\frac{2i\pi}{d})$ and $\varphi^l \;(l=0,1,\cdots,
d-1)$ are the settings of the appropriate phase shifters, for
convenience we denote them as a $d$ dimensional vector
$\vec{\varphi}=(\varphi^0,\varphi^1, \varphi^2,\cdots,
\varphi^{d-1})$. For state $|\psi_d^3\rangle$ of three-qudit
systems, the quantum prediction for the probabilities of obtaining
the outcome $(m, n, l)$ is then given by:
\begin{widetext}
\begin{eqnarray}\label{probablity}
P(X_1^{[i]}&=&m,X_2^{[j]}=n,X_3^{[k]}=l)=|\langle
mnl|U(\vec{\varphi}_{X_1^{[i]}})\otimes
U(\vec{\varphi}_{X_2^{[j]}})\otimes U(\vec{\varphi}_{X_3^{[k]}})|\psi_d^3\rangle|^2\nonumber\\
&=&\texttt{Tr}[(U^{\dagger}(\vec{\varphi}_{X_1^{[i]}})\otimes
U^{\dagger}(\vec{\varphi}_{X_2^{[j]}})\otimes
U^{\dagger}(\vec{\varphi}_{X_3^{[k]}}))|mnl\rangle\langle
mnl|(U(\vec{\varphi}_{X_1^{[i]}})\otimes
U(\vec{\varphi}_{X_2^{[j]}})\otimes
U(\vec{\varphi}_{X_3^{[k]}}))|\psi_d^3\rangle\langle\psi_d^3|].
\end{eqnarray}
\end{widetext}
Substituting Eq.~(\ref{probablity}) in to the
inequality~(\ref{three-BI}), one get the expression of $I^{[3]}_d$
in quantum mechanics. For the generalized GHZ state of three qubits:
\begin{eqnarray}\label{3qubits-GHZ}
|\psi_2^3\rangle=\cos\theta|000\rangle+\sin\theta|111\rangle,
\end{eqnarray}
numerical results show that when we set $\theta=\pi/4$,
$\vec{\varphi}_{X^{[1]}_1}=(0,-\pi/12)$,
$\vec{\varphi}_{X^{[2]}_1}=(0,\pi/4)$,
$\vec{\varphi}_{X^{[1]}_2}=(0,-\pi/6)$,
$\vec{\varphi}_{X^{[2]}_2}=(0,\pi/3)$,
$\vec{\varphi}_{X^{[1]}_3}=(0,0)$,
$\vec{\varphi}_{X^{[2]}_3}=(0,\pi/6)$, we get the maximal violation
$2\sqrt{2}$, which is the same of the maximal violation of CHSH
inequality for two qubits. For $\theta\in(0,\pi/8]$, the state
~(\ref{3qubits-GHZ}) doest not violate the inequality.
To measure the strength of violation of local realistic theories, we
may consider the mixed state
$\rho(F)=(1-F)|\psi_2^3\rangle\langle\psi_2^3|+\frac{F}{8}I\otimes
I\otimes I$, where $F$ ($0\leq F \leq1$) is the amount of the noise
present in the system~\cite{D.Kaszlikowski-2000} and $I$ is a
$2\times2$ identity matrix. 
According to the proposal introduced in
Ref.~\cite{D.Kaszlikowski-2000}, 
there exists some threshold value of $F$, denoted by $F_{thr}$, such
that for every $F\leq F_{thr}$, local and realistic description does
not exist. For inequality ~(\ref{three-BI}), the threshold fidelity
is $0.29289$, which is smaller than $1/2$,  the threshold fidelity
for MABK inequality for three qubits. This indicate that our
inequality is not as strong as the MABK inequality. Another set  of
states considered are the generalized W states:
$|\psi_2^3\rangle_W=\sin\beta\sin\xi|001\rangle+\sin\beta\cos\xi|010\rangle+\cos\beta|100\rangle$.
The maximal violation of this set of states is also $2\sqrt{2}$.
This result is surprising since for the previous inequalities, the
violations of the generalized W states are always smaller than that
of the GHZ states. Moreover, inequality~(\ref{three-BI}) is relevant
to three-qubit MABK inequality, i.e., there exist states which
violate inequality~(\ref{three-BI}) but do not violate the MABK
inequality. For instance, one may check that the state:
$|\Psi\rangle=0.169414|000\rangle+0.0461131|100\rangle+0.161369|101\rangle+
0.193624|110\rangle+0.951652|111\rangle$ do not violate the MABK
inequality but it do violate inequality~(\ref{three-BI}), and the
violation is $2.00382$.

For the generalized GHZ state of three qutrits:
\begin{eqnarray}
|\psi_3^3\rangle=\sin\theta_1\sin\theta_2|000\rangle+\sin\theta_1\cos\theta_2|111\rangle+\cos\theta_1|222\rangle\nonumber,
\end{eqnarray}
numerical results shows that when we set $\theta_1=0.9066$,
$\theta_2=0.6663$, $\vec{\varphi}_{X^{[1]}_1}=(0,-\pi/5,\pi/24)$,
$\vec{\varphi}_{X^{[2]}_1}=(0,\pi/24,-5\pi/12)$,
$\vec{\varphi}_{X^{[1]}_2}=(0,0,\pi/12)$,
$\vec{\varphi}_{X^{[2]}_2}=(0,\pi/3,-\pi/4)$,
$\vec{\varphi}_{X^{[1]}_3}=(0,\pi/30,\pi/20)$,
$\vec{\varphi}_{X^{[2]}_3}=(0,\pi/8,\pi/6)$, we get the maximal
violation $2.915$, which is the same of the maximal violation of the
CGLMP inequality for two qutrits. On the other hand, for the maximal
entangled state for three qutrits, namely $\theta_2=\pi/4$,
$\theta_1=\arccos(1/\sqrt{3})$, the quantum violation is $2.873$,
which is smaller than $2.915$. This indicts that the maximal
violation of inequality (\ref{three-BI}) occurs at the nonmaximally
entangled state. For higher dimensions, our numerical results show
that the maximal violation is similar to the CGLMP inequality and
the inequality (\ref{three-BI}) is also relevant to the inequalities
presented in Ref.\cite{Chen-3qudits-BI}.

The Bell inequalities can be easily generalized for arbitrary
$N$-qudit systems. The correlation functions in this case are in the
following form:
\begin{eqnarray}
Q_{i_1,\cdots,i_N}&=&\frac{1}{S}\sum_{x_1^{[i_1]}=0}^{d-1}\cdots\sum_{x_N^{[i_N]}=0}^{d-1}f^{i_1\cdots
i_j}(x_1^{[i_1]},\cdots,x_N^{[i_N]})\nonumber\\
&&\times P(X_1^{[i_1]}=x_1^{[i_1]},\cdots,
X_1^{[i_N]}=x_1^{[i_N]}),\nonumber
\end{eqnarray}
where  $S=(d-1)/2$, $f^{i_1\cdots
i_j}(x_1^{[i_1]},\cdots,x_N^{[i_N]})=S-M[(-1)^{\chi}(\sum_{j=1}^Nx_j^{[i_j]}),d]$,
which is similar to the definition of three-qudit correlation
functions and $\chi=\prod_{j=1}^Ni_j$. Based on these correlation
functions, the tight Bell inequality can be written as:
\begin{eqnarray}\label{N-qudits-BI}
I^{[2N]}_{d}=Q_{1\cdots1}+Q_{1212\cdots12}+Q_{2121\cdots21}-Q_{2\cdots2}\leq2,\;\;\;\;\\
I^{[2N+1]}_{d}=Q_{1\cdots1}+Q_{1212\cdots21}+Q_{2121\cdots12}-Q_{2\cdots2}\leq2.\nonumber
\end{eqnarray}
Using the same method as for the case of three qudits, one may check
that the the above inequalities~(\ref{N-qudits-BI}) are valid for
local hidden variable theory and they are tight.

For instance, we give two tight Bell inequalities. The first example
is the tight Bell inequality for four qudits:
\begin{eqnarray}\label{four-BI}
I^{[4]}_d=Q_{1111}+Q_{1212}+Q_{2121}-Q_{2222}\leq2,
\end{eqnarray}
Numerical results show that when $d=2$, the
inequality~(\ref{four-BI}) is maximally violated by the maximally
entangled state:
$|\psi_2^4\rangle=\frac{1}{\sqrt{2}}\left(|0000\rangle+|1111\rangle\right)$
if we set $\vec{\varphi}_{X^{[1]}_1}=(0,\pi/24)$,
$\vec{\varphi}_{X^{[2]}_1}=(0,\pi/12)$,
$\vec{\varphi}_{X^{[1]}_2}=(0,-\pi/6)$,
$\vec{\varphi}_{X^{[2]}_2}=(0,\pi/3)$,
$\vec{\varphi}_{X^{[1]}_3}=(0,-\pi/8)$,
$\vec{\varphi}_{X^{[2]}_3}=(0,\pi/3)$,
$\vec{\varphi}_{X^{[1]}_4}=(0,0)$, and
$\vec{\varphi}_{X^{[2]}_4}=(0,0)$. The violation is $2\sqrt{2}$,
which is the same as that of inequality~(\ref{three-BI}). Another
example is the  tight Bell inequality for five qudits:
\begin{eqnarray}\label{five-BI}
I^{[5]}_d=Q_{11111}+Q_{12121}+Q_{21212}-Q_{22222}\leq2.
\end{eqnarray}
Numerical results show that when $d=2$, the
inequality~(\ref{five-BI}) is maximally violated by the maximally
entangled state:
$|\psi_2^5\rangle=\frac{1}{\sqrt{2}}\left(|00000\rangle+|11111\rangle\right)$
when we set $\vec{\varphi}_{X^{[1]}_1}=(0,-\pi/12)$,
$\vec{\varphi}_{X^{[2]}_1}=(0,\pi/3)$,
$\vec{\varphi}_{X^{[1]}_2}=(0,-\pi/6)$,
$\vec{\varphi}_{X^{[2]}_2}=(0,\pi/3)$,
$\vec{\varphi}_{X^{[1]}_3}=(0,0)$,
$\vec{\varphi}_{X^{[2]}_3}=(0,\pi/12)$,
$\vec{\varphi}_{X^{[1]}_4}=(0,0)$,
$\vec{\varphi}_{X^{[2]}_4}=(0,0)$,
$\vec{\varphi}_{X^{[1]}_5}=(0,0)$, and
$\vec{\varphi}_{X^{[2]}_5}=(0,0)$. The violation is also
$2\sqrt{2}$. From the quantum violations of
inequalities~(\ref{three-BI}), (\ref{four-BI}) and (\ref{five-BI}),
we find that, different from the MABK, the quantum violations of our
inequalities remain the same, rather than increase,  with  the
increasing number of particles.

In summary, we have presented generic tight Bell inequalities for
arbitrary $N$-qudit systems based on the generalized correlation
functions. The new inequalities take the same simple form as the
CHSH inequality and when $N=2$ they reduce to the well known CGLMP
inequality. The new inequalities are not as strong as the MABK
inequality and  there exist some pure states that do not violate
these inequalities, while they are the first tight general Bell
inequalities for arbitrary $N$-qudit systems and they are relevant
to the previous known Bell inequalities. Frankly speaking, we do not
have a general proof of the tightness of these new inequalities.
Indeed, we have only checked that for small systems (namely,  three
qudits for $d\leq10$, four qudits for $d\leq7$, and five qudits for
$d\leq5$). Unfortunately, we have to leave this as an open question
here and we shall investigate it subsequently. Since the various use
of Bell inequality in quantum information, our results may be very
useful for the study of other Bell inequalities, quantum
entanglement measurement, distillation protocols, etc.

This work was supported in part by NSF of China (Grant No.
10605013), Program for New Century Excellent Talents in University,
and the Project-sponsored by SRF for ROCS, SEM.

\end{document}